\documentclass[aps,a4,reprint,twoside,floatfix,superscriptaddress]{revtex4-1}
\usepackage{graphicx}
\usepackage{bm}
\usepackage{bm}
\usepackage{color}

\begin{document}
\title[]{Enhancement of domain-wall mobility detected by NMR at the angular momentum compensation temperature}
\author{Masaki Imai}
\affiliation{Advanced Science Research Center, Japan Atomic Energy Agency, Tokai 319-1195, Japan}
\author{Hiroyuki Chudo}
\affiliation{Advanced Science Research Center, Japan Atomic Energy Agency, Tokai 319-1195, Japan}
\author{Mamoru Matsuo}
\affiliation{Advanced Science Research Center, Japan Atomic Energy Agency, Tokai 319-1195, Japan}
\affiliation{Riken Center for Emergent Matter Science (CEMS), Wako 351-0198, Japan}
\affiliation{Kavli Institute for Theoretical Sciences, University of Chinese Academy of Sciences,19 Yuquan Road, Beijing 100049, P.R.China}
\author{Sadamichi Maekawa}
\affiliation{Advanced Science Research Center, Japan Atomic Energy Agency, Tokai 319-1195, Japan}
\affiliation{Riken Center for Emergent Matter Science (CEMS), Wako 351-0198, Japan}
\affiliation{Kavli Institute for Theoretical Sciences, University of Chinese Academy of Sciences,19 Yuquan Road, Beijing 100049, P.R.China}
\author{Eiji Saitoh}
\affiliation{Advanced Science Research Center, Japan Atomic Energy Agency, Tokai 319-1195, Japan}
\affiliation{Advanced Institute for Materials Research, Tohoku University, Sendai 980-8577, Japan}
\affiliation{Institute for Materials Research, Tohoku University, Sendai 980-8577, Japan}
\affiliation{Department of Applied Physics, The University of Tokyo, Hongo, Bunkyo-ku, Tokyo, 113-8656, Japan}
\date{\today}

\begin{abstract}
The angular momentum compensation temperature $T_{\rm A}$ of ferrimagnets has attracted much attention because of high-speed magnetic dynamics near $T_{\rm A}$.
We show that NMR can be used to investigate domain wall dynamics near $T_{\rm A}$ in ferrimagnets.
We performed $^{57}$Fe-NMR measurements on the ferrimagnet Ho$_3$Fe$_5$O$_{12}$ with $T_{\rm A} = 245$~K.
In a multi-domain state, the NMR signal is enhanced by domain wall motion.
We found that the NMR signal enhancement shows a maximum at  $T_{\rm A}$ in the multi-domain state.
The NMR signal enhancement occurs due to increasing domain-wall mobility toward $T_{\rm A}$. We develop the NMR signal enhancement model involves domain-wall mobility. 
Our study shows that NMR in multi-domain state is a powerful tool to determine $T_{\rm A}$, even from a powder sample and it expands the possibility of searching for angular momentum-compensated materials.
\end{abstract}
\maketitle
\section{Introduction}
The angular momentum compensation in ferrimagnets, where angular momenta on different sublattices cancel each other out, has attracted much attention because of its unique character\cite{Stanciu2006,Stanciu2007,Binder2006,Kim2017,Imai2018,Zhifeng2018}.
In terms of an angular momentum, ferrimagnets at an angular momentum compensation temperature, $T_{\rm A}$, can be regarded as antiferromagnets, even though they have spontaneous magnetization.
Magnetic dynamics in ferrimagnets at $T_{\rm A}$ is also antiferromagnetic and much faster than in ferromagnets.
In ferrimagnetic resonance (FMR), for example, the Gilbert damping constant was predicted to be divergent at $T_{\rm A}$ \cite{Wangsness1953}.
The resonance frequency of the uniform mode arising from a ferromagnetic character increases and merges with that of an exchange mode arising from an antiferromagnetic character at $T_{\rm A}$ \cite{McGuire1955,Stanciu2006}, and the Gilbert damping parameter estimated from the linewidth of the uniform mode shows an anomaly near $T_{\rm A}$ \cite{Stanciu2006}.
Due to this fast magnetic dynamics, the high-speed magnetization reversal was realized in the amorphous ferrimagnet of GdFeCo alloy at $T_{\rm A}$ \cite{Stanciu2006}.

Moreover, in GdFeCo alloy, the domain wall mobility is enhanced at $T_{\rm A}$ \cite{Kim2017}.
Domain wall motion occurs due to the reorientation of magnetic moments.
Angular momentum accompanied by a magnetic moment prevents the magnetic moment from changing its direction due to the inertia of the angular momentum, and the domain-wall mobility is suppressed.
At $T_{\rm A}$, however, magnetic moments can easily change their direction because of the lack of inertia.
As a result, the domain-wall mobility is enhanced.
Thus, the angular momentum compensation of ferrimagnets may be useful for next-generation high-speed magnetic memories, such as racetrack memories\cite{parkin2008}.

The rare-earth iron garnet $R_3$Fe$_5$O$_{12}$ ($R$IG, where $R$ is a rare-earth element) is a ferrimagnet accompanied by $T_{\rm A}$ \cite{McGuire1955,LeCraw1965,Borghese1980}.
However, $R$IG does not show any anomaly in FMR at $T_{\rm A}$ because the angular momentum of $R^{3+}$ ions weakly couples with that of Fe$^{3+}$ ions and behaves almost as a free magnetic moment.
As a result, the magnetic relaxation frequency of the magnetic moment of $R^{3+}$ ions is much higher than that of the magnetic moment of Fe$^{3+}$ ions or the exchange frequency between $R^{3+}$ and Fe$^{3+}$ ions \cite{Rodrigue1960,Ohta1977,Srivastava1985,Kittel1959}.
In this case, the magnetic moment of $R^{3+}$ ions adiabatically follows the motion of the magnetic moment of Fe$^{3+}$ ions.
Hence, $R^{3+}$ ions contribute to the magnetization but not to the angular momentum due to heavy damping of the $R^{3+}$ site \cite{Kittel1959}.

Although $T_{\rm A}$ in $R$IG cannot be determined using FMR, the mobility of the bubble domains formed in the epitaxial thin film of the substituted $R$IG increases at a certain temperature, which is regarded as $T_{\rm A}$ \cite{Randoshkin2003}.
Furthermore, recently, it has become possible to directly and exactly measure the net angular momentum regardless of the material and its shape by using the Barnett effect, in which magnetization is induced by mechanical rotation due to spin--rotation coupling, \textcolor{black}{$H_{\rm SR} =-{\mathbf J} \cdot {\mathbf \Omega}$, where ${\mathbf J}$ and ${\mathbf \Omega}$ are the angular momentum of an electron and the angular velocity of the rotation \cite{Imai2018,Imai2019}.
When a sample is rotated, angular momenta of electrons in a magnetic material align along with the rotational axis, and, then, the material is magnetized without any external magnetic fields.
In this method, $T_{\rm A}$ is determined as the temperature where magnetization induced by the mechanical rotation vanishes because of the disappearance of the net angular momentum.}
Consequently, $T_{\rm A}$ of Ho$_3$Fe$_5$O$_{12}$ (HoIG) was determined to be 245~K \cite{Imai2018}.
With the focus on magnetic dynamics at $T_{\rm A}$, a microscopic method was required to investigate the spin dynamics at $T_{\rm A}$ regardless of materials and their shape.

\begin{figure}
\includegraphics[width=0.8\linewidth]{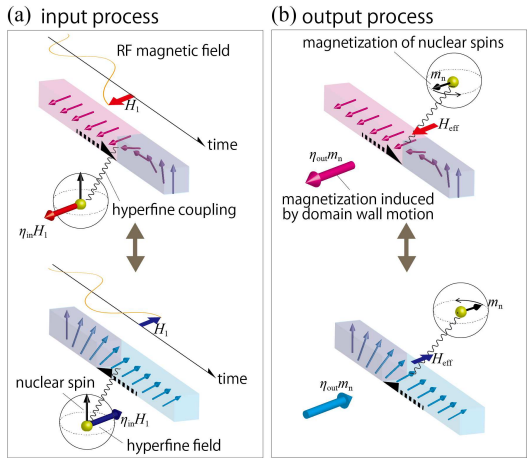}
\caption{Schematic illustration of enhancement of the NMR signal in a domain wall.
(a) An input RF magnetic field $H_1$ causes the domain wall to move.
The electron spins in the domain wall rotate, exciting the nuclear resonance through the hyperfine coupling.
As a result, $H_1$ appears to be $\eta_{\rm in}$ times for the nuclear spins.
(b) The domain wall moves in accordance with the precession of nuclear spins, and the bulk magnetization oscillates with NMR frequency.
The NMR signal becomes $\eta_{\rm out}$ times.
\label{fig1}}
\end{figure}

Here, we propose an NMR method to explore the spin dynamics at $T_{\rm A}$.
In a magnetic ordered state such as in ferromagnets and ferrimagnets, an NMR signal can be observed without any external magnetic field due to an internal field, which enables us to observe domain walls at zero or low magnetic fields.
Furthermore, the macroscopic magnetization of electrons enhances the NMR signal via hyperfine interactions.
Particularly, the NMR signal from nuclei in domain walls is strongly enhanced due to the magnetic domain wall motion, as shown in Fig.~\ref{fig1}.
An input radio frequency (RF) magnetic field $H_1$ used for NMR can move domain walls, thereby rotating magnetic moments in the walls and generating the transverse component of a hyperfine field in synchronization with the RF field.
As a result, $H_1$ is enhanced to become $\eta_{\rm in}H_1$, where $\eta_{\rm in}$ is the enhancement factor for the input process.
In the reverse process, the Larmor precession of nuclear spins causes domain wall motion, because the electronic system feels an effective magnetic field $H_{\rm eff}$ from the nuclear magnetization through the hyperfine interaction, which leads to the oscillation of the bulk magnetization; thus, a much stronger voltage is induced in the NMR pickup coil than the precession of nuclear magnetic moment $m_{\rm n}$, and the output NMR signal is enhanced to be $\eta_{\rm out}m_{\rm n}$, where $\eta_{\rm out}$ is the enhancement factor for the output process.
This enhancement effect enables us to selectively observe the NMR signal from nuclei in domain walls, even though the volume fraction of domain walls is much smaller than that of domains.

In this paper, we report results of an NMR study of HoIG under magnetic fields of up to 1.0~T.
For a multi-domain state below 0.3~T, the temperature dependence of the NMR intensity shows a maximum at $T_{\rm A}$.
On the other hand, for a single-domain state above 0.5~T, the temperature dependence of the NMR intensity does not show any anomalies at $T_{\rm A}$.
These results indicate the enhancement of the domain wall mobility at $T_{\rm A}$.
Extending a simple conventional model for describing $\eta_{\rm out}$\cite{Portis1960}, we formulated the modified enhancement factor $\eta'_{\rm out}$ by taking the domain wall mobility into account.
This enhancement of the NMR intensity at $T_{\rm A}$ enables us to estimate the domain wall mobility to determine $T_{\rm A}$, even in a powder sample.

\section{Experimental method}
We synthesized HoIG by solid-state reaction for this study\cite{Imai2018,Imai2019}. 
We ground the sample in a mortar to create a fine powder with a typical particle diameter of 5 $\mu$m.
The sample was packed in the NMR coil, which was perpendicular to the external magnetic field.
The NMR measurements of $^{57}$Fe nuclei at the $d$ site in Ho$_3$Fe$_5$O$_{12}$ were carried out using a standard phase-coherent pulsed spectrometer.
The NMR signals were obtained using the spin-echo method, with the first and second pulse durations of 1.0 and 2.0 $\mu {\rm s}$, respectively.
During the measurements, the pulse width was kept constant and the RF power was varied to maximize the NMR signal.
The spin-echo decay time $T_2$ was measured by varying the interval time $\tau$ between the first and second pulses.
The value of $T_2$ is defined such that $I(2\tau)=I(0)\exp(-2\tau/T_2)$, where $I(2\tau)$ and $I(0)$ are the NMR intensity at $2\tau$ and $\tau=0$, respectively.
The nuclear spin-lattice relaxation time $T_1$ was measured using the inversion recovery method.

\begin{figure*}
\includegraphics[width=\linewidth]{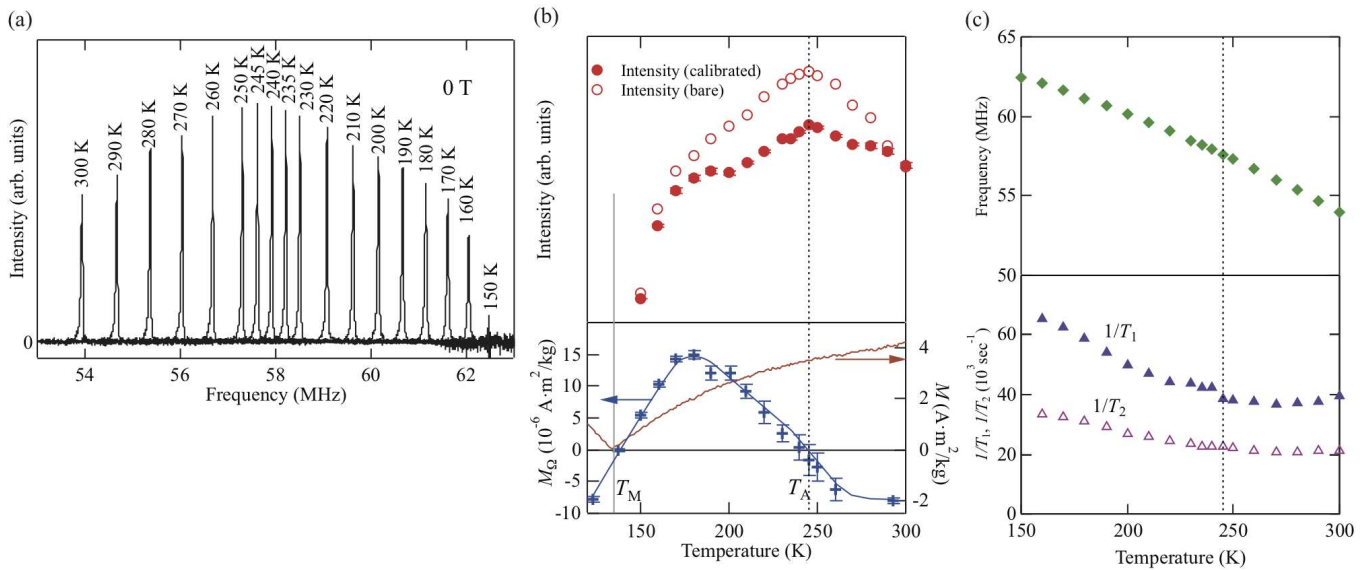}
\caption{The $^{57}$Fe NMR results for the $d$ site and the magnetic properties in Ho$_3$Fe$_5$O$_{12}$.
(a) Temperature dependence of the NMR spectra.
(b) In the upper panel, the red open and filled circles show the bare integrated signal intensity and calibrated intensity by multiplying by $T\nu^{-2}\exp(2\tau/T_2)$, respectively.
In the bottom panel, the blue cross shows $M_\Omega$ obtained by the Barnett effect.
The blue curve is a guide to the eye.
The orange curve shows the magnetization obtained under the magnetic field of 1000 [Oe].
In both panels, the black solid and dashed lines show the magnetization and angular momentum compensation temperatures of Ho$_3$Fe$_5$O$_{12}$, respectively.
(c) Temperature dependence of resonance frequency (top), $1/T_1$, and $1/T_2$ (bottom).\label{fig2}}
\end{figure*}

\section{Results}
Figure~\ref{fig2}(a) shows the temperature variation in the NMR spectra of $^{57}$Fe at the $d$ site without external fields.
Each NMR spectrum shows a single peak, and the peak shifts to higher frequencies with decreasing temperature.
The NMR intensity shows the maximum at 245~K.
The top panel of Fig.~\ref{fig2}(b) shows integrated NMR intensities.
Generally, the NMR intensities need to be calibrated when comparing them under different conditions.
The NMR intensity $I$ is proportional to the voltage induced in a pickup NMR coil by the precession of the nuclear magnetization $m_{\rm n}$.
Thus, $I$ is proportional to $dm_{\rm n}(t)/dt$.
Because $m_{\rm n}(t)$ rotates at the Larmor frequency $\nu$, $I$ is proportional to $\nu m_{\rm n}$.
The size of $m_{\rm n}$ depends on the polarization of the nuclear spin derived from the Boltzmann distribution function.
Thus $m_{\rm n}$ is proportional to $\nu/T$, where $T$ is the temperature.
As a result, $I$ is proportional to $\nu^2/T$.
Moreover, the NMR intensity measured by the spin echo method depends on $T_2$.
Therefore, we calibrated the NMR intensity by multiplying $T\nu^{-2}\exp(2\tau/T_2)$.

The calibrated NMR intensity is retained to show a maximum at 245~K, which coincides with $T_{\rm A}$ determined by the Barnett effect in which mechanical rotation induces magnetization $M_\Omega$ due to spin--rotation coupling \cite{Imai2018}.
The blue cross in the bottom panel of Fig.~\ref{fig2}(b) shows the temperature dependence of $M_\Omega$ under a rotation of 1500 Hz without any external magnetic field.
$M_\Omega$ becomes zero at two temperatures: The lower temperature coincides with \textcolor{black}{the magnetization compensation temperature} $T_{\rm M}$ determined by a conventional magnetization measurement \textcolor{black}{as shown by the orange curve in the bottom panel of Fig.~\ref{fig2}(b).}
At $T_{\rm M}$, spin--rotation coupling is effective, but $M_\Omega$ becomes zero due to the disappearance of bulk magnetization.
In contrast, the higher temperature can be assigned to $T_{\rm A}$, where the bulk magnetization remains but the spin--rotation coupling is not effective due to the disappearance of the net angular momentum \cite{Imai2018}.
Unlike the temperature dependence of the NMR intensity, there are no anomalies in the temperature dependence of $\nu$, $1/T_1$, and $1/T_2$ as shown in Fig.~\ref{fig2}(c).
These results indicate that the maximum NMR intensity can be attributed to an anomaly in the enhancement factor.

\begin{figure}
\includegraphics[width=1\linewidth]{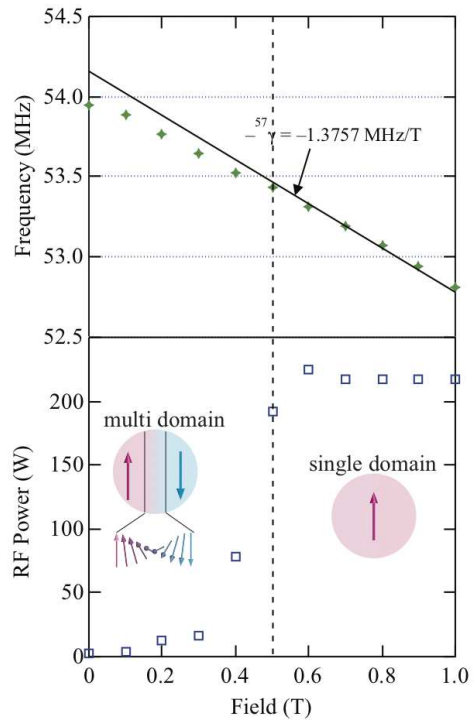}
\caption{The NMR results in the magnetic fields ranging from 0 to 1~T at 300~K.
The top panel shows the field dependence of the resonance frequency.
The solid line shows the slope of the gyromagnetic ratio of a $^{57}$Fe nucleus.
The bottom panel shows the field dependence of optimized RF power. \label{fig3}}
\end{figure}

To perform NMR experiments for the single-domain state, we characterized the magnetic field dependence of HoIG as shown in Fig.~\ref{fig3}. 
The top panel of Fig.~\ref{fig3} shows the NMR frequency in magnetic fields ranging from 0 to 1~T at 300~K.
With the increase in the magnetic field the resonance frequency decreases because the magnetic moment at the $d$ site aligns with the magnetic field above $T_{\rm M}$, and the hyperfine coupling constant is negative.
The line in the top panel of Fig.~\ref{fig3} shows a slope of $-^{57}\gamma =-1.3757$ MHz/T.
In the multi-domain state at low fields, the rate of decrease in the NMR frequency by applying external field is smaller than $-^{57}\gamma$ until all the domain walls disappear because the external field at nuclear positions is canceled out by the demagnetizing field caused by domain wall displacement due to the external magnetic field\cite{Yasuoka1964}. 
In the single-domain state above 0.6 T, the NMR frequency decreases with the ratio of $^{57}\gamma $ by the magnetic field.

The optimized RF input power is shown in the bottom panel of Fig.~\ref{fig3}.
At low magnetic fields, the RF input power is small due to the large $\eta_{\rm in}$, suggesting that the NMR signal from the domain walls, which is more enhanced than that from domains, dominates the NMR intensity.
The input power sharply increases in the region between 0.4 and 0.5~T and saturates above 0.6~T.
This result indicates that the domain structure changes from multi-domain to single domain between 0.4 and 0.5~T.
This is consistent with the result of the field dependence of the NMR frequency.
\textcolor{black}{At high magnetic fields, the NMR signal from the domain dominates the NMR intensity.}

\begin{figure}
\begin{center}
\includegraphics[width=1\linewidth]{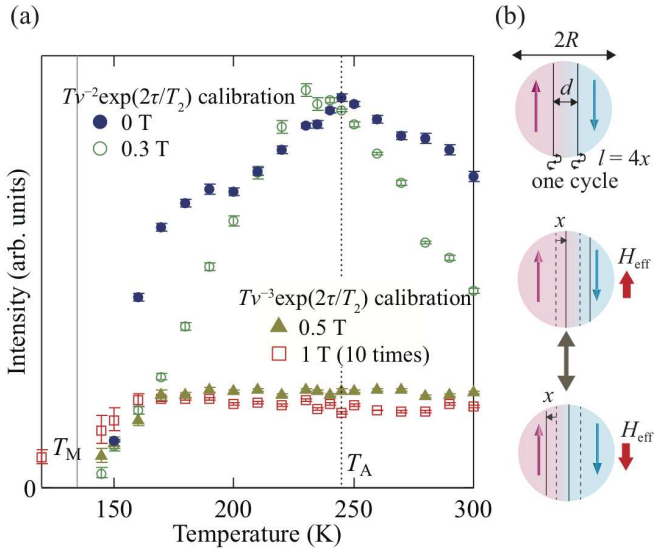}
\caption{(a) The temperature dependence of the NMR intensity in the magnetic fields ranging from 0 to 1~T.
The NMR intensity of 0 and 0.3 T is calibrated by multiplying by $T\nu^{-2}\exp(2\tau/T_2)$.
The NMR intensity of 0.5 and 1 T is calibrated by multiplying by $T\nu^{-3}\exp(2\tau/T_2)$.
The solid and dashed lines show the magnetization and angular momentum compensation temperatures, respectively.
(b) Schematic illustration of the domain wall motion induced by an effective RF magnetic field $H_{\rm eff}$ through the hyperfine coupling.
Here, $R$, $d$ and $x$ are a particle radius, a domain wall thickness, and domain wall displacement, respectively.
The domain wall moves $l=4x$ in one cycle of $t=1/\nu$.
\label{fig4}}
\end{center}
\end{figure}

Figure~\ref{fig4}(a) shows the temperature dependence of the calibrated NMR intensity in various magnetic fields.
In the multi-domain state at 0 and 0.3~T, the NMR intensity shows a maximum at 245 K and then decreases toward $T_{\rm M}$.
On the other hand, the calibrated values of the NMR intensity at 0.5 and 1.0~T do not show any anomalies around $T_{\rm A}$.
These results indicate that the maximum NMR intensity is attributed to the domain walls.
The drop in the NMR intensity at various magnetic fields around $T_{\rm M}$ results from the decrease in signal enhancement, which is proportional to the magnetization.
Notably, in the ferromagnetic or ferrimagnetic state, the enhancement factor $\eta_{\rm out}$ is proportional to the hyperfine field $H_{\rm n}$, which is also proportional to the NMR frequency $\nu$ \cite{Portis1960,Yasuoka1964}. 
Therefore, the NMR intensity in the single-domain state above 0.5 T is calibrated by multiplying by $T\nu^{-3}\exp(2\tau/T_2)$.
In the multi-domain state of this sample below 0.3~T, however, the enhancement factor does not depend on $\nu$ so that the NMR intensity below 0.3 T is calibrated by multiplying by $T\nu^{-2}\exp(2\tau/T_2)$.

The temperature at which the NMR intensity shows a maximum at 0.3~T decrease slightly.
It is speculated that $T_{\rm A}$ decreases under magnetic fields in $R$IG, because the expectation value of the angular momentum of $R^{3+}$ decreases above $T_{\rm M}$ in a magnetic field due to the decrease in molecular field at the $R$ site \cite{Imai2019}.

\section{discussion}
First, we introduce the conventional model describing NMR enhancement due to domain-wall motion \cite{Portis1960}.
In this model, the domain wall displacement $x$ is limited by a demagnetizing field $H_d(x)$.
The maximum displacement $x_{\rm max}$ is determined from a position in which $H_d(x_{\rm max})$ is balanced with an oscillating effective field $H_{\rm eff}$, which is created by the precession of the nuclear magnetic moment through the hyperfine interaction, $H_d(x_{\rm max})=H_{\rm eff}$.
Because the sample used for the NMR measurement in the present study is a powder, each particle in it is assumed to be spherical with radius $R$, as shown in Fig.~\ref{fig4}(b).
$H_d$ is expressed as $H_d(x_{\rm max})=2\pi \frac{x_{\rm max}}{R}m$.
The net electron magnetic moment $m$ is tilted by the effective field of $H_{\rm eff}=\frac{m_{\rm n}}{m}H_{\rm n}$, where $H_{\rm n}$ is the hyperfine field, and the tilt angle $\theta$ of $m$ can be described by $\theta=\pi x/d$, where $d$ is the domain-wall thickness.
Then, the bulk magnetization induced by nuclear magnetization is expressed as
\begin{equation}
\frac{RH_{\rm n}m_{\rm n}}{2dm} = \eta_{\rm out}m_{\rm n},  \label{eq1}
\end{equation}
where $\eta_{\rm out}$ is the enhancement factor of the NMR signal for the output process and is defined as $\eta_{\rm out}=\frac{RH_{\rm n}}{2dm}$.
This model assumes that the velocity of the domain-wall motion $v$ is fast enough to move $4x_{\rm max}$ during one cycle of oscillating effective field, i.e. $v=\mu H_{\rm eff}> 4x_{\rm max}\nu$, where $\mu$ is the domain wall mobility.

The conventional derivation of NMR enhancement induced by domain wall motion does not include the mobility of the domain-wall.
Herein, we consider that $v$ is not fast enough to follow the oscillating effective field, i.e., $v<4x_{\rm max}\nu$.
In this case, the displacement $x$ is limited by $\mu$.
Then, $x$ is expressed to be $v/4\nu$.
The enhancement factor $\eta_{\rm out}$ is modified such that
\begin{equation}
\eta'_{\rm out}=\frac{v}{4x_{\rm max}\nu}\eta_{\rm out} = \frac{\pi}{4d\gamma}\mu. \label{eq2} 
\end{equation}
This formula indicates that $\eta'_{\rm out}$ in the slow limit of domain wall motion is proportional to $\mu$, and $\eta'_{\rm out}$ in the fast limit of domain-wall motion is continually connected to the conventional $\eta_{\rm out}$. 
It is noted that $\eta'_{\rm out}$ does not depend on the NMR frequency $\nu$.

The domain-wall mobility of HoIG has not been reported, but it can be estimated from the reported damping parameters \cite{Vella-Coleiro1971,Vella‐Coleiro1972}.
The domain wall mobility of Gd$_3$Fe$_5$O$_{12}$ (GdIG) is 225 ${\rm m\cdot sec^{-1}Oe^{-1}}$ at 298~K \cite{Vella-Coleiro1971}.
The magnitude of the damping is inversely proportional to the domain wall mobility because the damping parameter of HoIG is 80 times as great as that of GdIG \cite{Vella‐Coleiro1972}, the domain wall mobility of HoIG at room temperature is estimated to be 2.8 ${\rm m\cdot sec^{-1}Oe^{-1}}$.
However, the domain-wall mobility required for motion $x_{\rm max}$ is defined such that $4x_{\rm max}\nu/H_{\rm eff}=2R\nu/\pi M$, which is estimated to be 4 $\rm {m\cdot sec^{-1}Oe^{-1}}$ for $4\pi M \sim 500$ G, $R \sim 5~\mu {\rm m}$, and $\nu \sim 50$~MHz.
Thus, this evaluation indicates that, in HoIG, the displacement of domain walls induced by nuclear precession is limited by $\mu$.
Therefore, in the multi-domain state in HoIG, we used the modified enhancement factor $\eta'_{\rm out}$ in Eq.~(\ref{eq2}).
We estimate the value of $\mu$ at $T_{\rm A}$ to be 3.5 ${\rm m\cdot sec^{-1}Oe^{-1}}$ using $\mu=2.8$ ${\rm m\cdot sec^{-1}Oe^{-1}}$ at 300 K.
When we assume $d$ to be 0.1--1.0 $\mu$m, $\eta_{\rm out}$ is estimated to be 10$^2$--10$^3$, which is comparable to typical enhancement factors \cite{Portis1960,Dho1997}.
Thus, the NMR method is very sensitive to detect such a small enhancements at $T_{\rm A}$.

\begin{acknowledgments}
We thank H. Yasuoka for fruitful discussion.
This work was supported by JST ERATO Grant Number JPMJER1402, JSPS Grant-in-Aid for Scientific Research on Innovative Areas Grant Number JP26103005, and JSPS KAKENHI Grant Numbers JP16H04023, JP17H02927.
\end{acknowledgments}
M. I and H. C contributed equally to this work.


\begin{thebibliography}{22}%
\makeatletter
\providecommand \@ifxundefined [1]{%
 \@ifx{#1\undefined}
}%
\providecommand \@ifnum [1]{%
 \ifnum #1\expandafter \@firstoftwo
 \else \expandafter \@secondoftwo
 \fi
}%
\providecommand \@ifx [1]{%
 \ifx #1\expandafter \@firstoftwo
 \else \expandafter \@secondoftwo
 \fi
}%
\providecommand \natexlab [1]{#1}%
\providecommand \enquote  [1]{``#1''}%
\providecommand \bibnamefont  [1]{#1}%
\providecommand \bibfnamefont [1]{#1}%
\providecommand \citenamefont [1]{#1}%
\providecommand \href@noop [0]{\@secondoftwo}%
\providecommand \href [0]{\begingroup \@sanitize@url \@href}%
\providecommand \@href[1]{\@@startlink{#1}\@@href}%
\providecommand \@@href[1]{\endgroup#1\@@endlink}%
\providecommand \@sanitize@url [0]{\catcode `\\12\catcode `\$12\catcode
  `\&12\catcode `\#12\catcode `\^12\catcode `\_12\catcode `\%12\relax}%
\providecommand \@@startlink[1]{}%
\providecommand \@@endlink[0]{}%
\providecommand \url  [0]{\begingroup\@sanitize@url \@url }%
\providecommand \@url [1]{\endgroup\@href {#1}{\urlprefix }}%
\providecommand \urlprefix  [0]{URL }%
\providecommand \Eprint [0]{\href }%
\providecommand \doibase [0]{http://dx.doi.org/}%
\providecommand \selectlanguage [0]{\@gobble}%
\providecommand \bibinfo  [0]{\@secondoftwo}%
\providecommand \bibfield  [0]{\@secondoftwo}%
\providecommand \translation [1]{[#1]}%
\providecommand \BibitemOpen [0]{}%
\providecommand \bibitemStop [0]{}%
\providecommand \bibitemNoStop [0]{.\EOS\space}%
\providecommand \EOS [0]{\spacefactor3000\relax}%
\providecommand \BibitemShut  [1]{\csname bibitem#1\endcsname}%
\let\auto@bib@innerbib\@empty
\bibitem [{\citenamefont {Stanciu}\ \emph {et~al.}(2006)\citenamefont
  {Stanciu}, \citenamefont {Kimel}, \citenamefont {Hansteen}, \citenamefont
  {Tsukamoto}, \citenamefont {Itoh}, \citenamefont {Kirilyuk},\ and\
  \citenamefont {Rasing}}]{Stanciu2006}%
  \BibitemOpen
  \bibfield  {author} {\bibinfo {author} {\bibfnamefont {C.~D.}\ \bibnamefont
  {Stanciu}}, \bibinfo {author} {\bibfnamefont {A.~V.}\ \bibnamefont {Kimel}},
  \bibinfo {author} {\bibfnamefont {F.}~\bibnamefont {Hansteen}}, \bibinfo
  {author} {\bibfnamefont {A.}~\bibnamefont {Tsukamoto}}, \bibinfo {author}
  {\bibfnamefont {A.}~\bibnamefont {Itoh}}, \bibinfo {author} {\bibfnamefont
  {A.}~\bibnamefont {Kirilyuk}}, \ and\ \bibinfo {author} {\bibfnamefont
  {T.}~\bibnamefont {Rasing}},\ }\href {\doibase 10.1103/PhysRevB.73.220402}
  {\bibfield  {journal} {\bibinfo  {journal} {Phys. Rev. B}\ }\textbf {\bibinfo
  {volume} {73}},\ \bibinfo {pages} {220402} (\bibinfo {year}
  {2006})}\BibitemShut {NoStop}%
\bibitem [{\citenamefont {Stanciu}\ \emph {et~al.}(2007)\citenamefont
  {Stanciu}, \citenamefont {Tsukamoto}, \citenamefont {Kimel}, \citenamefont
  {Hansteen}, \citenamefont {Kirilyuk}, \citenamefont {Itoh},\ and\
  \citenamefont {Rasing}}]{Stanciu2007}%
  \BibitemOpen
  \bibfield  {author} {\bibinfo {author} {\bibfnamefont {C.~D.}\ \bibnamefont
  {Stanciu}}, \bibinfo {author} {\bibfnamefont {A.}~\bibnamefont {Tsukamoto}},
  \bibinfo {author} {\bibfnamefont {A.~V.}\ \bibnamefont {Kimel}}, \bibinfo
  {author} {\bibfnamefont {F.}~\bibnamefont {Hansteen}}, \bibinfo {author}
  {\bibfnamefont {A.}~\bibnamefont {Kirilyuk}}, \bibinfo {author}
  {\bibfnamefont {A.}~\bibnamefont {Itoh}}, \ and\ \bibinfo {author}
  {\bibfnamefont {T.}~\bibnamefont {Rasing}},\ }\href {\doibase
  10.1103/PhysRevLett.99.217204} {\bibfield  {journal} {\bibinfo  {journal}
  {Phys. Rev. Lett.}\ }\textbf {\bibinfo {volume} {99}},\ \bibinfo {pages}
  {217204} (\bibinfo {year} {2007})}\BibitemShut {NoStop}%
\bibitem [{\citenamefont {Binder}\ \emph {et~al.}(2006)\citenamefont {Binder},
  \citenamefont {Weber}, \citenamefont {Mosendz}, \citenamefont {Woltersdorf},
  \citenamefont {Izquierdo}, \citenamefont {Neudecker}, \citenamefont {Dahn},
  \citenamefont {Hatchard}, \citenamefont {Thiele}, \citenamefont {Back},\ and\
  \citenamefont {Scheinfein}}]{Binder2006}%
  \BibitemOpen
  \bibfield  {author} {\bibinfo {author} {\bibfnamefont {M.}~\bibnamefont
  {Binder}}, \bibinfo {author} {\bibfnamefont {A.}~\bibnamefont {Weber}},
  \bibinfo {author} {\bibfnamefont {O.}~\bibnamefont {Mosendz}}, \bibinfo
  {author} {\bibfnamefont {G.}~\bibnamefont {Woltersdorf}}, \bibinfo {author}
  {\bibfnamefont {M.}~\bibnamefont {Izquierdo}}, \bibinfo {author}
  {\bibfnamefont {I.}~\bibnamefont {Neudecker}}, \bibinfo {author}
  {\bibfnamefont {J.~R.}\ \bibnamefont {Dahn}}, \bibinfo {author}
  {\bibfnamefont {T.~D.}\ \bibnamefont {Hatchard}}, \bibinfo {author}
  {\bibfnamefont {J.-U.}\ \bibnamefont {Thiele}}, \bibinfo {author}
  {\bibfnamefont {C.~H.}\ \bibnamefont {Back}}, \ and\ \bibinfo {author}
  {\bibfnamefont {M.~R.}\ \bibnamefont {Scheinfein}},\ }\href {\doibase
  10.1103/PhysRevB.74.134404} {\bibfield  {journal} {\bibinfo  {journal} {Phys.
  Rev. B}\ }\textbf {\bibinfo {volume} {74}},\ \bibinfo {pages} {134404}
  (\bibinfo {year} {2006})}\BibitemShut {NoStop}%
\bibitem [{\citenamefont {Kim}\ \emph {et~al.}(2017)\citenamefont {Kim},
  \citenamefont {Kim}, \citenamefont {Hirata}, \citenamefont {Oh},
  \citenamefont {Tono}, \citenamefont {Kim}, \citenamefont {Okuno},
  \citenamefont {Ham}, \citenamefont {Kim}, \citenamefont {Go} \emph
  {et~al.}}]{Kim2017}%
  \BibitemOpen
  \bibfield  {author} {\bibinfo {author} {\bibfnamefont {K.-J.}\ \bibnamefont
  {Kim}}, \bibinfo {author} {\bibfnamefont {S.~K.}\ \bibnamefont {Kim}},
  \bibinfo {author} {\bibfnamefont {Y.}~\bibnamefont {Hirata}}, \bibinfo
  {author} {\bibfnamefont {S.-H.}\ \bibnamefont {Oh}}, \bibinfo {author}
  {\bibfnamefont {T.}~\bibnamefont {Tono}}, \bibinfo {author} {\bibfnamefont
  {D.-H.}\ \bibnamefont {Kim}}, \bibinfo {author} {\bibfnamefont
  {T.}~\bibnamefont {Okuno}}, \bibinfo {author} {\bibfnamefont {W.~S.}\
  \bibnamefont {Ham}}, \bibinfo {author} {\bibfnamefont {S.}~\bibnamefont
  {Kim}}, \bibinfo {author} {\bibfnamefont {G.}~\bibnamefont {Go}},  \emph
  {et~al.},\ }\href {\doibase doi:10.1038/nmat4990} {\bibfield  {journal}
  {\bibinfo  {journal} {Nature materials}\ }\textbf {\bibinfo {volume} {16}},\
  \bibinfo {pages} {1187} (\bibinfo {year} {2017})}\BibitemShut {NoStop}%
\bibitem [{\citenamefont {Imai}\ \emph {et~al.}(2018)\citenamefont {Imai},
  \citenamefont {Ogata}, \citenamefont {Chudo}, \citenamefont {Ono},
  \citenamefont {Harii}, \citenamefont {Matsuo}, \citenamefont {Ohnuma},
  \citenamefont {Maekawa},\ and\ \citenamefont {Saitoh}}]{Imai2018}%
  \BibitemOpen
  \bibfield  {author} {\bibinfo {author} {\bibfnamefont {M.}~\bibnamefont
  {Imai}}, \bibinfo {author} {\bibfnamefont {Y.}~\bibnamefont {Ogata}},
  \bibinfo {author} {\bibfnamefont {H.}~\bibnamefont {Chudo}}, \bibinfo
  {author} {\bibfnamefont {M.}~\bibnamefont {Ono}}, \bibinfo {author}
  {\bibfnamefont {K.}~\bibnamefont {Harii}}, \bibinfo {author} {\bibfnamefont
  {M.}~\bibnamefont {Matsuo}}, \bibinfo {author} {\bibfnamefont
  {Y.}~\bibnamefont {Ohnuma}}, \bibinfo {author} {\bibfnamefont
  {S.}~\bibnamefont {Maekawa}}, \ and\ \bibinfo {author} {\bibfnamefont
  {E.}~\bibnamefont {Saitoh}},\ }\href {\doibase 10.1063/1.5041464} {\bibfield
  {journal} {\bibinfo  {journal} {Appl. Phys. Lett.}\ }\textbf {\bibinfo
  {volume} {113}},\ \bibinfo {pages} {052402} (\bibinfo {year} {2018})},\
  \Eprint {http://arxiv.org/abs/https://doi.org/10.1063/1.5041464}
  {https://doi.org/10.1063/1.5041464} \BibitemShut {NoStop}%
\bibitem [{\citenamefont {Zhu}\ \emph {et~al.}(2018)\citenamefont {Zhu},
  \citenamefont {Fong},\ and\ \citenamefont {Liang}}]{Zhifeng2018}%
  \BibitemOpen
  \bibfield  {author} {\bibinfo {author} {\bibfnamefont {Z.}~\bibnamefont
  {Zhu}}, \bibinfo {author} {\bibfnamefont {X.}~\bibnamefont {Fong}}, \ and\
  \bibinfo {author} {\bibfnamefont {G.}~\bibnamefont {Liang}},\ }\href
  {\doibase 10.1103/PhysRevB.97.184410} {\bibfield  {journal} {\bibinfo
  {journal} {Phys. Rev. B}\ }\textbf {\bibinfo {volume} {97}},\ \bibinfo
  {pages} {184410} (\bibinfo {year} {2018})}\BibitemShut {NoStop}%
\bibitem [{\citenamefont {Wangsness}(1953)}]{Wangsness1953}%
  \BibitemOpen
  \bibfield  {author} {\bibinfo {author} {\bibfnamefont {R.~K.}\ \bibnamefont
  {Wangsness}},\ }\href {\doibase 10.1103/PhysRev.91.1085} {\bibfield
  {journal} {\bibinfo  {journal} {Phys. Rev.}\ }\textbf {\bibinfo {volume}
  {91}},\ \bibinfo {pages} {1085} (\bibinfo {year} {1953})}\BibitemShut
  {NoStop}%
\bibitem [{\citenamefont {McGuire}(1955)}]{McGuire1955}%
  \BibitemOpen
  \bibfield  {author} {\bibinfo {author} {\bibfnamefont {T.~R.}\ \bibnamefont
  {McGuire}},\ }\href {\doibase 10.1103/PhysRev.97.831.2} {\bibfield  {journal}
  {\bibinfo  {journal} {Phys. Rev.}\ }\textbf {\bibinfo {volume} {97}},\
  \bibinfo {pages} {831} (\bibinfo {year} {1955})}\BibitemShut {NoStop}%
\bibitem [{\citenamefont {Parkin}\ \emph {et~al.}(2008)\citenamefont {Parkin},
  \citenamefont {Hayashi},\ and\ \citenamefont {Thomas}}]{parkin2008}%
  \BibitemOpen
  \bibfield  {author} {\bibinfo {author} {\bibfnamefont {S.~S.}\ \bibnamefont
  {Parkin}}, \bibinfo {author} {\bibfnamefont {M.}~\bibnamefont {Hayashi}}, \
  and\ \bibinfo {author} {\bibfnamefont {L.}~\bibnamefont {Thomas}},\
  }\href@noop {} {\bibfield  {journal} {\bibinfo  {journal} {Science}\ }\textbf
  {\bibinfo {volume} {320}},\ \bibinfo {pages} {190} (\bibinfo {year}
  {2008})}\BibitemShut {NoStop}%
\bibitem [{\citenamefont {LeCraw}\ \emph {et~al.}(1965)\citenamefont {LeCraw},
  \citenamefont {Remeika},\ and\ \citenamefont {Matthews}}]{LeCraw1965}%
  \BibitemOpen
  \bibfield  {author} {\bibinfo {author} {\bibfnamefont {R.~C.}\ \bibnamefont
  {LeCraw}}, \bibinfo {author} {\bibfnamefont {J.~P.}\ \bibnamefont {Remeika}},
  \ and\ \bibinfo {author} {\bibfnamefont {H.}~\bibnamefont {Matthews}},\
  }\href {\doibase 10.1063/1.1714259} {\bibfield  {journal} {\bibinfo
  {journal} {J. Appl. Phys.}\ }\textbf {\bibinfo {volume} {36}},\ \bibinfo
  {pages} {901} (\bibinfo {year} {1965})},\ \Eprint
  {http://arxiv.org/abs/https://doi.org/10.1063/1.1714259}
  {https://doi.org/10.1063/1.1714259} \BibitemShut {NoStop}%
\bibitem [{\citenamefont {Borghese}\ \emph {et~al.}(1980)\citenamefont
  {Borghese}, \citenamefont {Cosmi}, \citenamefont {De~Gasperis},\ and\
  \citenamefont {Tappa}}]{Borghese1980}%
  \BibitemOpen
  \bibfield  {author} {\bibinfo {author} {\bibfnamefont {C.}~\bibnamefont
  {Borghese}}, \bibinfo {author} {\bibfnamefont {R.}~\bibnamefont {Cosmi}},
  \bibinfo {author} {\bibfnamefont {P.}~\bibnamefont {De~Gasperis}}, \ and\
  \bibinfo {author} {\bibfnamefont {R.}~\bibnamefont {Tappa}},\ }\href
  {\doibase 10.1103/PhysRevB.21.183} {\bibfield  {journal} {\bibinfo  {journal}
  {Phys. Rev. B}\ }\textbf {\bibinfo {volume} {21}},\ \bibinfo {pages} {183}
  (\bibinfo {year} {1980})}\BibitemShut {NoStop}%
\bibitem [{\citenamefont {Rodrigue}\ \emph {et~al.}(1960)\citenamefont
  {Rodrigue}, \citenamefont {Meyer},\ and\ \citenamefont
  {Jones}}]{Rodrigue1960}%
  \BibitemOpen
  \bibfield  {author} {\bibinfo {author} {\bibfnamefont {G.~P.}\ \bibnamefont
  {Rodrigue}}, \bibinfo {author} {\bibfnamefont {H.}~\bibnamefont {Meyer}}, \
  and\ \bibinfo {author} {\bibfnamefont {R.~V.}\ \bibnamefont {Jones}},\ }\href
  {\doibase 10.1063/1.1984756} {\bibfield  {journal} {\bibinfo  {journal} {J.
  Appl. Phys.}\ }\textbf {\bibinfo {volume} {31}},\ \bibinfo {pages} {S376}
  (\bibinfo {year} {1960})},\ \Eprint
  {http://arxiv.org/abs/https://doi.org/10.1063/1.1984756}
  {https://doi.org/10.1063/1.1984756} \BibitemShut {NoStop}%
\bibitem [{\citenamefont {Ohta}\ \emph {et~al.}(1977)\citenamefont {Ohta},
  \citenamefont {Ikeda}, \citenamefont {Ishida},\ and\ \citenamefont
  {Sugita}}]{Ohta1977}%
  \BibitemOpen
  \bibfield  {author} {\bibinfo {author} {\bibfnamefont {N.}~\bibnamefont
  {Ohta}}, \bibinfo {author} {\bibfnamefont {T.}~\bibnamefont {Ikeda}},
  \bibinfo {author} {\bibfnamefont {F.}~\bibnamefont {Ishida}}, \ and\ \bibinfo
  {author} {\bibfnamefont {Y.}~\bibnamefont {Sugita}},\ }\href {\doibase
  10.1143/JPSJ.43.705} {\bibfield  {journal} {\bibinfo  {journal} {J. Phys.
  Soc. Jpn.}\ }\textbf {\bibinfo {volume} {43}},\ \bibinfo {pages} {705}
  (\bibinfo {year} {1977})},\ \Eprint
  {http://arxiv.org/abs/https://doi.org/10.1143/JPSJ.43.705}
  {https://doi.org/10.1143/JPSJ.43.705} \BibitemShut {NoStop}%
\bibitem [{\citenamefont {Srivastava}\ \emph {et~al.}(1985)\citenamefont
  {Srivastava}, \citenamefont {Uma Maheshwar~Rao},\ and\ \citenamefont
  {Hanumantha~Rao}}]{Srivastava1985}%
  \BibitemOpen
  \bibfield  {author} {\bibinfo {author} {\bibfnamefont {C.~M.}\ \bibnamefont
  {Srivastava}}, \bibinfo {author} {\bibfnamefont {B.}~\bibnamefont {Uma
  Maheshwar~Rao}}, \ and\ \bibinfo {author} {\bibfnamefont {N.~S.}\
  \bibnamefont {Hanumantha~Rao}},\ }\href {\doibase 10.1007/BF02747577}
  {\bibfield  {journal} {\bibinfo  {journal} {Bulletin of Materials Science}\
  }\textbf {\bibinfo {volume} {7}},\ \bibinfo {pages} {237} (\bibinfo {year}
  {1985})}\BibitemShut {NoStop}%
\bibitem [{\citenamefont {Kittel}(1959)}]{Kittel1959}%
  \BibitemOpen
  \bibfield  {author} {\bibinfo {author} {\bibfnamefont {C.}~\bibnamefont
  {Kittel}},\ }\href {\doibase 10.1103/PhysRev.115.1587} {\bibfield  {journal}
  {\bibinfo  {journal} {Phys. Rev.}\ }\textbf {\bibinfo {volume} {115}},\
  \bibinfo {pages} {1587} (\bibinfo {year} {1959})}\BibitemShut {NoStop}%
\bibitem [{\citenamefont {Randoshkin}\ \emph {et~al.}(2003)\citenamefont
  {Randoshkin}, \citenamefont {Polezhaev}, \citenamefont {Sysoev},\ and\
  \citenamefont {Sazhin}}]{Randoshkin2003}%
  \BibitemOpen
  \bibfield  {author} {\bibinfo {author} {\bibfnamefont {V.~V.}\ \bibnamefont
  {Randoshkin}}, \bibinfo {author} {\bibfnamefont {V.~A.}\ \bibnamefont
  {Polezhaev}}, \bibinfo {author} {\bibfnamefont {N.~N.}\ \bibnamefont
  {Sysoev}}, \ and\ \bibinfo {author} {\bibfnamefont {Y.~N.}\ \bibnamefont
  {Sazhin}},\ }\href {\doibase 10.1134/1.1562240} {\bibfield  {journal}
  {\bibinfo  {journal} {Physics of the Solid State}\ }\textbf {\bibinfo
  {volume} {45}},\ \bibinfo {pages} {513} (\bibinfo {year} {2003})}\BibitemShut
  {NoStop}%
\bibitem [{\citenamefont {Imai}\ \emph {et~al.}(2019)\citenamefont {Imai},
  \citenamefont {Chudo}, \citenamefont {Ono}, \citenamefont {Harii},
  \citenamefont {Matsuo}, \citenamefont {Ohnuma}, \citenamefont {Maekawa},\
  and\ \citenamefont {Saitoh}}]{Imai2019}%
  \BibitemOpen
  \bibfield  {author} {\bibinfo {author} {\bibfnamefont {M.}~\bibnamefont
  {Imai}}, \bibinfo {author} {\bibfnamefont {H.}~\bibnamefont {Chudo}},
  \bibinfo {author} {\bibfnamefont {M.}~\bibnamefont {Ono}}, \bibinfo {author}
  {\bibfnamefont {K.}~\bibnamefont {Harii}}, \bibinfo {author} {\bibfnamefont
  {M.}~\bibnamefont {Matsuo}}, \bibinfo {author} {\bibfnamefont
  {Y.}~\bibnamefont {Ohnuma}}, \bibinfo {author} {\bibfnamefont
  {S.}~\bibnamefont {Maekawa}}, \ and\ \bibinfo {author} {\bibfnamefont
  {E.}~\bibnamefont {Saitoh}},\ }\href {\doibase 10.1063/1.5095166} {\bibfield
  {journal} {\bibinfo  {journal} {Appl. Phys. Lett.}\ }\textbf {\bibinfo
  {volume} {114}},\ \bibinfo {pages} {162402} (\bibinfo {year} {2019})},\
  \Eprint {http://arxiv.org/abs/https://doi.org/10.1063/1.5095166}
  {https://doi.org/10.1063/1.5095166} \BibitemShut {NoStop}%
\bibitem [{\citenamefont {Portis}\ and\ \citenamefont
  {Gossard}(1960)}]{Portis1960}%
  \BibitemOpen
  \bibfield  {author} {\bibinfo {author} {\bibfnamefont {A.~M.}\ \bibnamefont
  {Portis}}\ and\ \bibinfo {author} {\bibfnamefont {A.~C.}\ \bibnamefont
  {Gossard}},\ }\href {\doibase 10.1063/1.1984666} {\bibfield  {journal}
  {\bibinfo  {journal} {J. Appl. Phys.}\ }\textbf {\bibinfo {volume} {31}},\
  \bibinfo {pages} {S205} (\bibinfo {year} {1960})},\ \Eprint
  {http://arxiv.org/abs/https://doi.org/10.1063/1.1984666}
  {https://doi.org/10.1063/1.1984666} \BibitemShut {NoStop}%
\bibitem [{\citenamefont {Yasuoka}(1964)}]{Yasuoka1964}%
  \BibitemOpen
  \bibfield  {author} {\bibinfo {author} {\bibfnamefont {H.}~\bibnamefont
  {Yasuoka}},\ }\href {\doibase 10.1143/JPSJ.19.1182} {\bibfield  {journal}
  {\bibinfo  {journal} {J. Phys. Soc. Jpn.}\ }\textbf {\bibinfo {volume}
  {19}},\ \bibinfo {pages} {1182} (\bibinfo {year} {1964})},\ \Eprint
  {http://arxiv.org/abs/https://doi.org/10.1143/JPSJ.19.1182}
  {https://doi.org/10.1143/JPSJ.19.1182} \BibitemShut {NoStop}%
\bibitem [{\citenamefont {{Vella-Coleiro}}\ \emph {et~al.}(1971)\citenamefont
  {{Vella-Coleiro}}, \citenamefont {{Smith}},\ and\ \citenamefont {{Van
  Uitert}}}]{Vella-Coleiro1971}%
  \BibitemOpen
  \bibfield  {author} {\bibinfo {author} {\bibfnamefont {G.}~\bibnamefont
  {{Vella-Coleiro}}}, \bibinfo {author} {\bibfnamefont {D.}~\bibnamefont
  {{Smith}}}, \ and\ \bibinfo {author} {\bibfnamefont {L.}~\bibnamefont {{Van
  Uitert}}},\ }\href {\doibase 10.1109/TMAG.1971.1067188} {\bibfield  {journal}
  {\bibinfo  {journal} {IEEE Transactions on Magnetics}\ }\textbf {\bibinfo
  {volume} {7}},\ \bibinfo {pages} {745} (\bibinfo {year} {1971})}\BibitemShut
  {NoStop}%
\bibitem [{\citenamefont {Vella‐Coleiro}\ \emph {et~al.}(1972)\citenamefont
  {Vella‐Coleiro}, \citenamefont {Smith},\ and\ \citenamefont
  {Van~Uitert}}]{Vella‐Coleiro1972}%
  \BibitemOpen
  \bibfield  {author} {\bibinfo {author} {\bibfnamefont {G.}~\bibnamefont
  {Vella‐Coleiro}}, \bibinfo {author} {\bibfnamefont {D.}~\bibnamefont
  {Smith}}, \ and\ \bibinfo {author} {\bibfnamefont {L.}~\bibnamefont
  {Van~Uitert}},\ }\href {\doibase 10.1063/1.1654209} {\bibfield  {journal}
  {\bibinfo  {journal} {Appl. Phys. Lett.}\ }\textbf {\bibinfo {volume} {21}},\
  \bibinfo {pages} {36} (\bibinfo {year} {1972})},\ \Eprint
  {http://arxiv.org/abs/https://doi.org/10.1063/1.1654209}
  {https://doi.org/10.1063/1.1654209} \BibitemShut {NoStop}%
\bibitem [{\citenamefont {Dho}\ \emph {et~al.}(1997)\citenamefont {Dho},
  \citenamefont {Kim}, \citenamefont {Lee},\ and\ \citenamefont
  {Lee}}]{Dho1997}%
  \BibitemOpen
  \bibfield  {author} {\bibinfo {author} {\bibfnamefont {J.}~\bibnamefont
  {Dho}}, \bibinfo {author} {\bibfnamefont {M.}~\bibnamefont {Kim}}, \bibinfo
  {author} {\bibfnamefont {S.}~\bibnamefont {Lee}}, \ and\ \bibinfo {author}
  {\bibfnamefont {W.-J.}\ \bibnamefont {Lee}},\ }\href {\doibase
  10.1063/1.363872} {\bibfield  {journal} {\bibinfo  {journal} {J. Appl.
  Phys.}\ }\textbf {\bibinfo {volume} {81}},\ \bibinfo {pages} {1362} (\bibinfo
  {year} {1997})},\ \Eprint
  {http://arxiv.org/abs/https://doi.org/10.1063/1.363872}
  {https://doi.org/10.1063/1.363872} \BibitemShut {NoStop}%
\end{thebibliography}
\end{document}